\documentclass[twocolumn,showpacs,pre]{revtex4}
\usepackage{dcolumn}
\usepackage{graphicx}
\usepackage{graphicx,amssymb,amsmath,revsymb}
\usepackage{natbib}

\begin{document}

\title{Linear surface roughness growth and flow smoothening in a three-dimensional biofilm model}

\author{D. A. Head$^{1}$}

\affiliation{$^{1}$School of Computing, University of Leeds, Leeds LS2 9JT, United Kingdom.}

\date{\today}

\begin{abstract}
The sessile microbial communities known as biofilms exhibit varying architectures as environmental factors are varied, which for immersed biofilms includes the shear rate of the surrounding flow. Here we modify an established agent-based biofilm model to include affine flow, and employ it to analyse the growth of surface roughness of single-species, three-dimensional biofilms. We find linear growth laws for surface geometry in both horizontal and vertical directions, and measure the thickness of the active surface layer, which is shown to anti-correlate with roughness. Flow is shown to monotonically reduce surface roughness without affecting the thickness of the active layer. We argue that the rapid roughening is due to non-local surface interactions mediated by the nutrient field, which are curtailed when advection competes with diffusion. We further argue the need for simplified models to elucidate the underlying mechanisms coupling flow to growth.
\end{abstract}



\maketitle

%
%
\section{Introduction}

Biofilms are surface-associated sessile microbial communities encased in a protective polymeric matrix at least partly of their own production~\cite{CostertonBook,CommunityBook}. Part of the healthy human microbiome~\cite{HMP,PhilBook}, they can also be deleterious when harbouring pathogenic species and protecting them from biocidal treatment, such as in water distribution systems or medical implants~\cite{DeirdreBook,Walker2004}. Biofilm architectures take a variety of forms, including flat, rough, rippled and columnar, depending on both environmental ({\em e.g.} shear flow, nutrient supply) and intrinsic ({\em e.g.} cell motility, intracellular communication) factors~\cite{Stoodley1999,Shrout2006,Barken2008}. Structure can affect function, such as the frequently-observed channels that are thought to permit nutrient penetration deep into the film~\cite{Wood2000}. A deep, quantitative understanding into the relationship between biofilm structure and flow would therefore suggest strategies for eradicating or otherwise modulating biofilm formation, but no theory with predictive capability currently exists.

The quantitative description of the growth of rough surfaces, both biotic and abiotic, is an established field in statistical physics, in particular when the surface geometry is scale-invariant or {\em fractal}~\cite{BarabasiBook}. Analytical and numerical treatments of model systems have demonstrated that their large length-scale behaviour can typically be grouped into a small number of so-called {\em universality classes}. Which class a specific system falls into depends on invariant intrinsic properties, such as dimension, symmetries and conserved quantities, and also the nature of the interactions between separated surface points, {\em i.e.} whether such interactions are {\em local} (strictly short-ranged) or {\em non-local}. In the latter case, growth at one surface point depends (in principle) on the current geometry of the entire surface. Such non-locality is known to drastically alter the fractal surface growth picture~\cite{BarabasiBook,Tang1990}.

Bacterial~\cite{Lacasta1999,Mimura2000} and fungal~\cite{Lopez1998,Lopez2002} colonies have been investigated within the framework of fractal surface growth~\cite{Paiva2007}. However, the relevance of these findings to biofilms, and to which (if any) universality class biofilms belong, remains unclear. A recent two-dimensional study employing a somewhat realistic model for biofilm growth found complex behaviour that could not be facilely interpreted using established paradigms~\cite{Bonachela2011}. In addition, none of the aforementioned models incorporate flow, despite the significant effect of biofilm architecture this is known to have~\cite{Stoodley1999}. Some models have been designed that do incorporate fluid-structure coupling, but not for growing films represented on the cellular scale as here: The model of Alpkvist and Klapper, which uses the immersed boundary method to couple the biomass to Navier-Stokes equations, does not include scalar fields or biofilm growth~\cite{Alpkvist2007}. Biofilm growth is also absent in the two-dimensional model of Picioreanu {\em et al.}~\cite{Picioreanu2000}, and this also represents the biofilm on the continuum level, which is inappropriate for studying cell-scale features. Other two-dimensional continuum models have also been developed~\cite{Towler2007,Duddu2009} which may be relevant at larger length scales than those considered here.

%
%
\begin{figure}[htbp]
\centerline{\includegraphics[width=8.5cm]{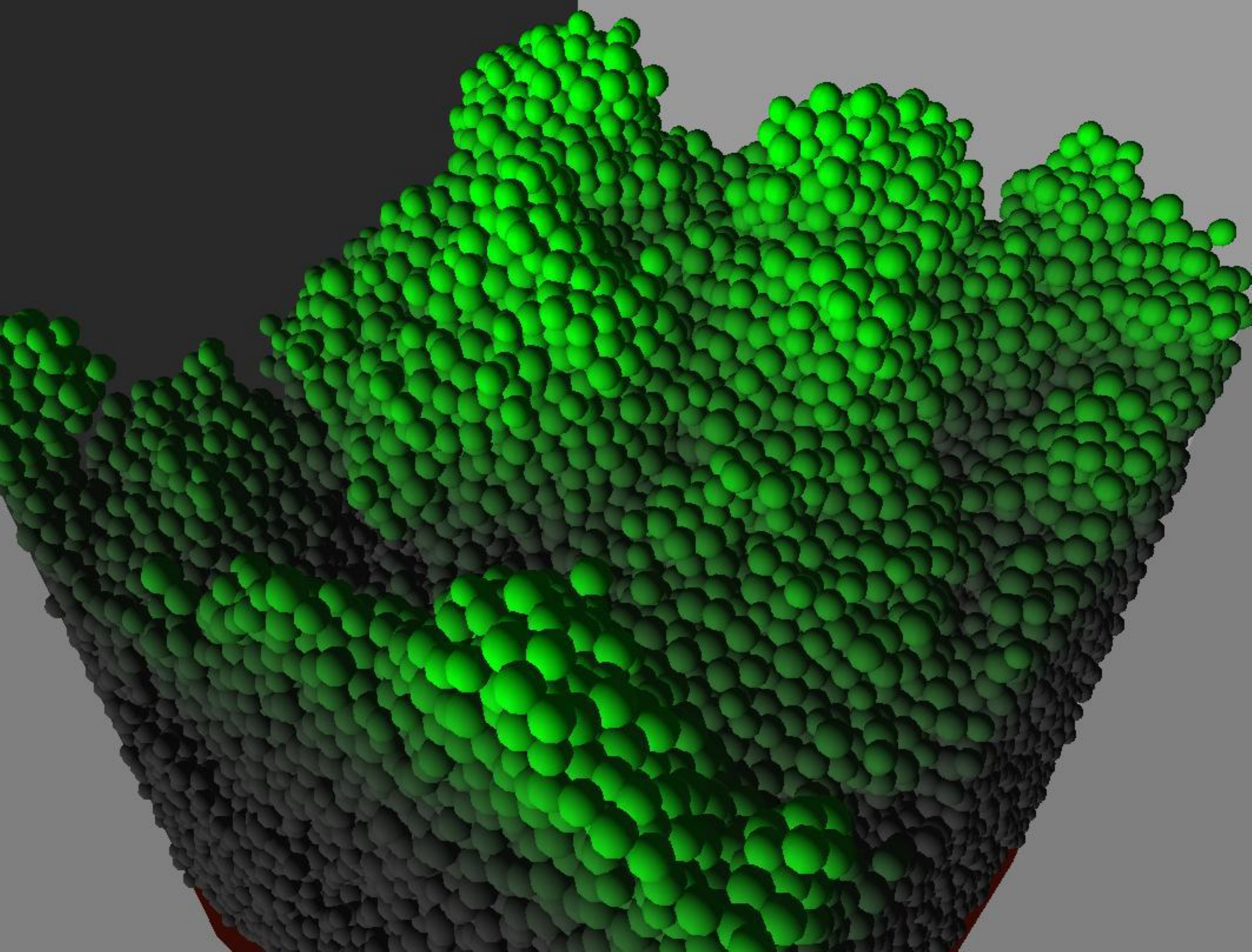}}
\caption{Snapshot of system state for $\dot{\gamma}=0$. Particle brightness is proportional to their metabolic reaction rates~$r_{i}$. The system size is $L_{x}=L_{y}=40\,d^{\rm max}$ and the bulk nutrient concentration is $c_{0}=10K_{1/2}$ (see text for details). A full color version with the nutrient field also displayed is available from the supplementary information~\cite{SuppInf}.}
\label{f:snapshot}
\end{figure}

In this article, we introduce an agent-based biofilm model in which both the nutrient field and the biofilm itself is coupled to the flow, and analyse it within the framework of fractal surface growth. Our model extends the {\em Individual-based Model} (IbM)~\cite{Lardon2011,Kreft2001,Wang2010,Xavier2005} by incorporating adhesive links between nearby particles, replacing the purely-repulsive `pushing' rules that such models typically employ. This small but far-reaching extension generates a mechanically consistent biofilm that can react to shear stresses applied by the flow. A snapshot of our model, which we dub the {\em mechanical IbM} model or {\em m-IbM}, is shown in Fig.~\ref{f:snapshot} and~\cite{SuppInf}. Analysis reveals a rapid growth of surface roughness, both parallel and normal to the direction of mean surface growth, that is far more rapid than the scaling obeyed by canonical models~\cite{BarabasiBook}, which we attribute to a non-local surface interaction deriving from the long-range effects of nutrient depletion. Switching on flow, we observe a similar growth law but with a smaller coefficient, corresponding to {\em smoother} biofilms. We argue this is due to the competition between nutrient diffusion and advection, and that high advection modulates the non-local surface interactions resulting in a less rough film.

This manuscript is arranged as follows. In Sec.~\ref{s:model} we detail the modules in our model, how they are coupled,  and the algorithms employed to iterate them during growth. In Sec.~\ref{s:results} we describe analysis of growing films in the presence of nutrient fields of varying concentrations, taking care to control the finite size effects that are ever-present in scale-invariant systems. Starting without flow, we quantify the growth of surface roughness parallel and perpendicular to the mean direction of growth using standard metrics, in both cases finding the aforementioned linear growth laws. We also relate the depth of actively-growing particles near the surface to the roughness, confirming previous findings~\cite{Nadell2010}. Repeating the analysis with flow reveals a systematic reduction in roughness as the flow rate increases, while not affecting the thickness of this active layer. In Sec.~\ref{s:discussion} we attempt to place our findings into the broader context of fractal surface growth. Two appendices are reserved for technical details. In Appendix~\ref{s:appThy} we derive analytical expressions for the growth of a flat film, which is used to compare to the numerical results. Finally, in Appendix~\ref{s:appData} we explain how the surface heights were extracted from our off-lattice simulations.

%
%
\begin{figure}[htbp]
\centerline{\includegraphics[width=8cm]{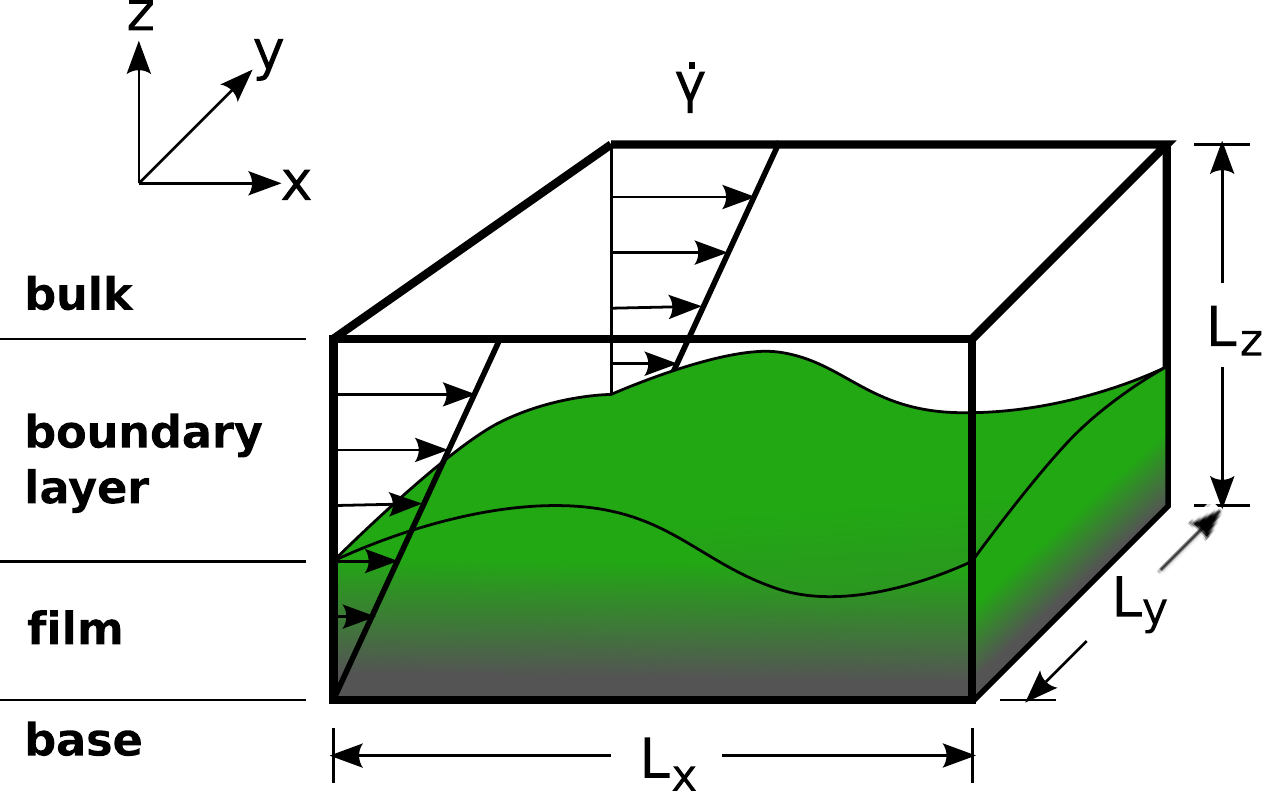}}
\caption{Schematic of the model. The simulation domain consists of the biofilm and the boundary layer, which lie between the solid base at $z=0$ and the bulk fluid at $z=L_{z}$. Periodic boundary conditions are assumed in the $x$ and $y$ directions. When flow is present, it takes the form of an affine shear parallel to the $x$-axis with fixed rate~$\dot{\gamma}$.}
\label{f:schematicDiagram}
\end{figure}

%
%
\section{Model definition}
\label{s:model}

The simulation model employed here is based on the {\em Individual-based Model} or IbM, which is an established agent-based method for the mathematical modelling of biofilms~\cite{Lardon2011,Kreft2001,Wang2010,Xavier2005}. This hybrid scheme couples discrete entities representing cells or cell aggregates to one or more continuous scalar fields, representing soluble factors such as nutrients or metabolites. In our scheme, we introduce a single vector field corresponding to the fluid velocity that couples to both the scalar fields and the biofilm, the latter through the requirement of mechanical stability as explained below. We also associate a mass of EPS (the {\em Extracellular Polymeric Substances} that make up the biofilm matrix~\cite{Flemming2010}) with each particle, and this is used to determine the elastic interactions between particles. We first present an overview of the central variables in each component of the model, before describing the time evolution of each in detail. A summary of the physical parameters and variables for each module is given in Table~\ref{t:varParam}.

\subsection{Variables and parameters}
\label{s:params}

Our model contains three components or modules, referred to as {\em biomass}, {\em scalars} and {\em fluid}, sharing the same spatial domain of a rectangular box with dimensions $(L_{x},L_{y},L_{z})$. See Fig.~\ref{f:schematicDiagram} for a schematic diagram of the system geometry. The solid surface to which the biofilm is attached corresponds to the $z=0$ plane, and the bulk fluid corresponds to the upper plane $z=L_{z}$. Fluid flow (if present) is parallel to the $x$-axis. Periodic boundary conditions are assumed in the $x$ and $y$ directions to avoid introducing wall or edge effects.

\begin{figure}[htbp]
\centerline{\includegraphics[width=8cm]{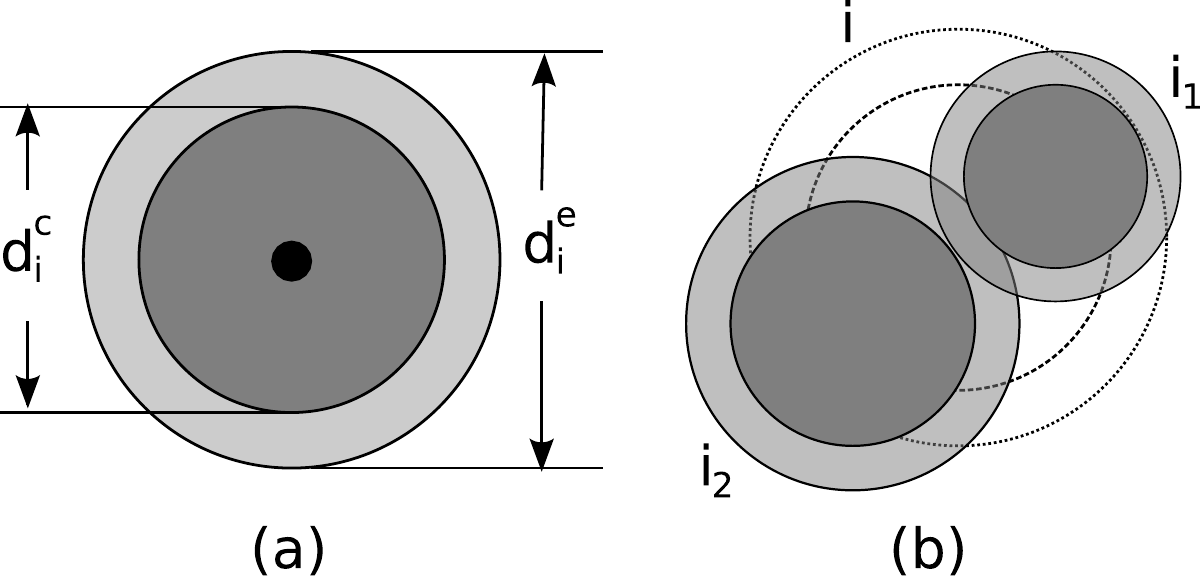}}
\caption{(a)~Single particle $i$ with cell diameter $d^{\rm c}_{i}$ and EPS diameter~$d^{\rm e}_{i}$. (b)~Schematic of redistribution of cellular and EPS masses after particle $i$ (dashed lines) divides into $i_{1}$ and $i_{2}$. Each mass component is conserved.}
\label{f:cellEPS}
\end{figure}

The {\em biomass} module consists of $N(t)$ biomass particles $i=1\ldots N(t)$ at time $t$, each with a cellular mass $m^{\rm c}_{i}$ and an EPS mass $m^{\rm e}_{i}$ (see Fig.~\ref{f:cellEPS}(a)). The centres of the particles are denoted ${\bf x}_{i}$. Each particle is regarded as spherical, with a cell diameter $d^{\rm c}_{i}$ that can be related to the common cell density $\rho^{\rm c}$ by $d^{\rm c}_{i}=\sqrt[3]{6m^{\rm c}_{i}/\pi\rho^{\rm c}}$. The EPS associated with particle $i$ is regarded as forming a spherical shell of density $\rho^{\rm e}$ extending from the cell surface. The outer diameter of this shell is denoted $d^{e}_{i}$ and is related to the EPS density by $d^{e}_{i}=\sqrt[3]{6m^{e}_{i}/\pi\rho^{\rm e}+(d^{\rm c}_{i})^{3}}$.

For this application, the {\em scalar} module consists of a single scalar field $c({\bf x})$ corresponding to the concentration of the soluble nutrient. This enters the system from the bulk as per the boundary condition $c(z=L_{z})=c_{0}$ (for simplicity, depletion and replenishment of $c_{0}$ with time is not considered). Cells reduce the local nutrient to fuel their increase in mass. This reaction is regarded as localised at the centre ${\bf x}_{i}$ of each particle~$i$, with a reaction rate given by the commonly-employed Michaelis-Menten form~\cite{HoggBook}, which includes the particle mass~$m_{i}$,
\begin{equation}
r_{i}=-k_{\rm max}m_{i}[1+K_{1/2}/c({\bf x}_{i})]^{-1}.
\label{e:monod}
\end{equation}
This form, in which a linear dependence on concentration crosses over to a saturated rate when $c\gg K_{1/2}$, is commonly employed for models in the IbM template~\cite{Kreft2001}. Metabolic activity is converted into an increase in both cellular and EPS masses as $\partial_{t}m^{c}_{i}=Y^{\rm c}|r_{i}|$ and $\partial_{t}m^{\rm e}_{i}=Y^{\rm e}_{\rm rel}\partial_{t}m^{\rm c}_{i}=Y^{\rm e}_{\rm rel}Y^{\rm c}|r_{i}|$.

Finally, the {\em fluid} module describes the fluid velocity field ${\bf v}({\bf x})$. Here only a simple affine shear flow is considered, {\em i.e.} ${\bf v}(x,y,z)=(\dot{\gamma}z,0,0)$ with $\dot{\gamma}$ the constant shear rate.

The initial state was taken to be a sub-monolayer of particles with number density $\rho^{\rm IC}$ per unit surface area. Particles were added at random uniformly over the surface, and attempted additions that would create particles with overlapping cell radii were rejected. Each seed particle was anchored to a point directly beneath it (see below for the definition of anchors).

Although the three model components share the same spatial domain, they relax on separated time scales, allowing them to be solved sequentially: The fluid iteration relaxes on times of the order of $ms$, the scalars on the order of $s$ and the biomass on the order of  $min$ to $hour$. The iteration cycle proceeds in the order {\em scalar} $\rightarrow$ {\em biomass} $\rightarrow$ {\em fluid} $\rightarrow$ {\em scalar} $\rightarrow$ \ldots, with data extraction just before the {\em biomass} growth iteration. Each stage in this cycle is now explained in detail.

%
%

%
%
\begin{table}
\begin{ruledtabular}
\begin{tabular}{clc}
Label & Meaning & Value \\
\hline
$c_{0}$ & Bulk nutrient concentration & - \\
$\dot{\gamma}$ & Fluid shear rate & - \\
$L_{x}$ & Box length in direction of flow & - \\
$L_{y}$ & Box width in vorticity direction & - \\
\hline
$d^{\rm max}$ & Division diameter & 5$\mu$m \\
$L_{z}$ & Height from solid surface to bulk & 80 $d^{\rm max}$ \\
$\rho^{c}$ & Cell density (excluding water) & 0.2 pg/$\mu$m$^{3}$ \\
$\rho^{e}$ & EPS density (excluding water) & 4$\times10^{-2}$ pg/$\mu$m$^{3}$ \\
$K_{1/2}$ & Reaction saturation concentration & $10^{-6}$ pg/$\mu$m$^{3}$ \\
$D$ & Nutrient diffusion coefficient & $10^{3}$ $\mu$m$^{2}$/s \\
$k_{\rm max}$ & Base reaction rate & 0.5/h \\
$Y^{\rm c}$ & Yield factor for cell mass & 0.2 \\
$Y^{\rm e}_{\rm rel}$ & Relative yield factor for EPS & 0.4 \\
$\sigma^{\rm div}$ & Width of relative mass division & 0.1 \\
$\nu$ & Fluid viscosity & $10^{-3}$ Pa s\\
$k^{\rm anc}$ & Anchor spring stiffness & 50 pN/$\mu$m \\
$\kappa^{\rm e}$ & EPS spring stiffness per mass & 5 pN $\mu$m$^{-1}$ pg$^{-1}$ \\
$\rho^{\rm IC}$ & Initial surface number density & $10^{-2}$ $\mu$m$^{-2}$ \\
\end{tabular}
\end{ruledtabular}
\caption{Variables and parameters. Those above the line are treated as variables here, while those below were kept fixed with the values quoted, which were chosen to be representative of oral bacteria taking simple sugars as a nutrient~\cite{Hamilton1979,Ellwood1982,Marsh1985}.}
\label{t:varParam}
\end{table}

\subsection{Scalar iteration}

The nutrient concentration $c({\bf x})$ obeys the steady-state reaction-diffusion-advection equation
\begin{equation}
0
=
\partial_{t}c
=
-{\bf v}\cdot\nabla c
+
D\nabla^{2}c
+
\sum_{i=1}^{N}r_{i}\delta({\bf x}-{\bf x}_{i})
\label{e:rda}
\end{equation}
obeying the mixed boundary conditions
\begin{eqnarray}
c(z=L_{z}) & = & c_{0}, \\
\left.\partial_{z}c\right|_{z=0} & = & 0.
\end{eqnarray}
That this can be solved separately to biofilm growth is a direct consequence of the separation of time scales mentioned in the previous paragraph. The reaction rates $r_{i}$ are given by~(\ref{e:monod}), and note that diffusion is assumed to be constant. This is solved numerically using a finite difference scheme solved on a regular rectangular mesh using geometric multigrid~\cite{BriggsBook}. To determine the reaction terms in (\ref{e:rda}), the value of $c$ at the particle centre ${\bf x}_{i}$ is found by trilinear interpolation from the adjacent mesh nodes, inserting into (\ref{e:monod}), and then distributing the resulting $r_{i}$ onto the same lattice nodes in a way that conserves the total reaction rate. Here we weight the contribution to each node by the inverse of its distance from~${\bf x}_{i}$.

\subsection{Biomass iteration}

Once the steady-state reaction rates $r_{i}$ have been determined for each particle~$i$, the increase in both cellular mass $m^{\rm c}_{i}$ and the EPS mass $m^{\rm e}_{i}$ are found by multiplying the mass growth rates by the biomass time interval~$\Delta t^{\rm bio}$. This time-step is adaptive, so that higher relative growth rates correspond to smaller time steps and {\em vice versa}. A linear variation was employed here, $\Delta t^{\rm bio}=C \max_{i=1\ldots N}\left(\frac{1}{m_{i}}\partial_{t}m_{i}\right)$, with $C=0.01$ to fix the maximum particle growth at around 1\% per time step. $C$ was varied to ensure no discernible variation of measured quantities. Note that this biomass growth time step is distinct from, and many orders of magnitude larger than, the time step used during fluid stabilisation described below.

After the cellular and EPS masses of each particle, and thus their corresponding diameters, have been updated, the system is checked for division events. Any particles whose new diameter exceeds the division threshold, {\em i.e.}  $d^{\rm c}_{i}>d^{\rm max}$, divides into two daughter cells $i_{1}$ and~$i_{2}$. Mass is conserved during division, but is distributed asymmetrically between the two daughters according to $m^{\rm c}_{i_1}=m^{\rm c}_{i}-m^{\rm c}_{i_2}=\lambda_{i}m^{\rm c}_{i}$, where $\lambda_{i}$ is a random variable chosen for each division event from a Normal distribution with mean $\frac{1}{2}$ and width~$\sigma^{\rm div}$. The EPS mass is divided similarly, with the same~$\lambda_{i}$. The daughter cells are placed at opposite poles of a sphere, centred on the parent cell, with a diameter $\frac{1}{2}(d^{\rm c}_{i_1}+d^{\rm e}_{i_1})+\frac{1}{2}(d^{\rm c}_{i_2}+d^{\rm e}_{i_2})$ so that their EPS shells overlap; see Fig.~\ref{f:cellEPS}(b). The axis of the sphere on which the daughters are added is chosen at random to ensure division cannot introduce anisotropy.

The links between the particles can now be determined. In essence, this amounts to identifying pairs of particles $i$ and $j$ whose EPS shells overlap, $|{\bf x}_{i}-{\bf x}_{j}|<\frac{1}{2}(d^{\rm e}_{i}+d^{\rm e}_{j})$, and adding a spring between their centres. In practice this leads to the rare instances where both daughter cells become disconnected from the film shortly after a division event. This is ultimately an artefact of the simplistic representation of the EPS as a spherical shell surrounding the particle - in a real biofilm, the EPS would deform during division to continuously enmesh both daughter particles. To robustly maintain film integrity, after each round of division events, all particles are sorted into clusters, where two particles belong to the same cluster if their EPS shells overlap. Any isolated clusters are translated into contact with either the film or the base at $z=0$, whichever requires the shortest motion. Note that such translations are always small, much less than particle diameters, and can (and typically do) include horizontal components, thus this does not introduce any form of smoothing. Furthermore, to avoid `knots' of springs, no particle is allowed to have more than 13 links. Any particle with more than this number of links has its longest links removed until this maximum number is reached. The actual maximum value does not measurably alter the results, unless it becomes very high; 13 was chosen as the maximum number of identical spheres that can touch a central one in a disordered packing.

Links between particle pairs are deleted before each growth and division cycle, and recreated afterwards. They are therefore transient links that reflect the current configuration of the film. Links to the base at $z=0$ are different in that they cannot move once formed, else the film could drift in an uncontrolled manner in the presence of flow. Instead, these {\em anchor} links are permanent and do not move once formed. They are created when a particle that does not already have an anchor link comes into contact with the base, {\em i.e.} has a centre at a height $z_{i}<\frac{1}{2}d^{\rm e}_{i}$. A spring is then created between the particle and an {\em anchor point} that is directly below the particle position at this time, {\em i.e.} at $(x_{i},y_{i},0)$. An anchor is not created if the particle already has 3 transient links to anchored particles. These rules maintain a stable population of anchor links that does not drift during the biofilm evolution.

\subsection{Fluid iteration}

In a full model with a spatio-temporally varying flow field, ${\bf v}({\bf x})$ would need to be simultaneously solved with the stabilisation of the biomass in a momentum-conserving manner. Since ${\bf v}({\bf x})$ is fixed here, we need only stabilise the film in the presence of flow. This amounts to demanding that the net force ${\bf f}_{i}$ on each particle $i$ simultaneously vanishes. Two forces contribute to ${\bf f}_{i}$, a {\em drag force} deriving from the flow, and a {\em matrix force} due to the links between particles, or anchor links to the base. The drag force is based on Stokes flow past a sphere,
\begin{equation}
{\bf f}_{i}^{\rm drag}=3\pi\nu d^{\rm c}_{i}{\bf v}({\bf x}_{i})
\end{equation}
where the fluid viscosity $\nu$ is chosen to be that of water. The matrix force ${\bf f}^{\rm mat}_{i}$ derives from the links determined in the previous step that are now identified as Hookean springs ({\em i.e.} linear springs that are repulsive when contracted and adhesive when stretched). For anchor links, the spring force is $k^{\rm anc}(r-\frac{1}{2}d^{\rm c}_{i})$ where $r$ is the distance of the cell centre from the anchor point on the surface, and $k^{\rm anc}$ is a uniform spring constant. This scalar force is projected along the line connecting the particle to the anchor point to give the required vector force. For the transient EPS-mediated links between particle pairs $i$ and~$j$, the force is $\kappa^{\rm e}m^{\rm e}_{ij}(|{\bf x}_{i}-{\bf x}_{j}|-\ell_0)$ with a natural length $\ell_0=\frac{1}{4}(d^{\rm c}_{i}+d^{\rm c}_{j}+d^{\rm e}_{i}+d^{\rm e}_{j})$ corresponding to the midpoint of the EPS shells. Here, $\kappa^{\rm e}$ is the stiffness per unit mass, and $m^{\rm e}_{ij}$ is the mass of the EPS that is attributed to this link. This is determined by equally distributing each particle's EPS mass to each of its (non-anchor) links. This scalar force is projected to the line of centres between ${\bf x}_{i}$ and ${\bf x}_{j}$ in an equal-and-opposite manner.

The goal is to determine the whole film configuration $\{{\bf x}_{i}\}$ for which each ${\bf f}_{i}={\bf f}^{\rm drag}_{i}+{\bf f}^{\rm mat}_{i}=0$. Two methods were used here which gave equivalent results. Since they are standard they will only be described briefly here. The {\em non-linear conjugate gradient method}~\cite{BonnansBook}, which was found to be most efficient for small systems, requires repeated construction and inversion of a large, sparse {\em stiffness matrix} giving the changes in each component of each force for small changes in particle positions. Block-diagonal preconditioning was also used. The second method, which proved to be more efficient for large systems and those with flow, was to use {\em overdamped molecular dynamics}~\cite{AllenTildesley} in which particles were moved in the direction of their unbalanced force: $\Delta{\bf x}_{i}=\Delta t \,{\bf f}_i/3\pi\nu d^{\rm c}_{i}$, where an adaptive time step $\Delta t$ was used that increases as the largest velocity decreases. For both methods, convergence tolerances were systematically varied until there was no discernible variation in measured quantities.

%
%
\section{Results}
\label{s:results}

The control variables are here chosen to be those that are also amenable to experimental control, namely the bulk nutrient concentration $c_{0}$ and the shear rate $\dot{\gamma}$. The horizontal surface dimensions $L_{x}=L_{y}$ are systematically varied to determine finite size effects. All other parameters are kept fixed with the values quoted in Table~\ref{t:varParam}, which were taken to be representative of oral bacteria growing in the presence of sugars~\cite{Hamilton1979,Ellwood1982,Marsh1985}. The theoretical predictions for flat films referred to below are derived in Appendix~\ref{s:appThy}. Unless otherwise stated, all results are presented in terms of dimensionless quantities constructed by scaling by combinations of the length $d^{\rm max}$, inverse time $k_{\rm max}$ and mass concentration~$K_{1/2}$.

\subsection{Surface roughening without flow}

\begin{figure}[htbp]
\center{\includegraphics[width=9cm]{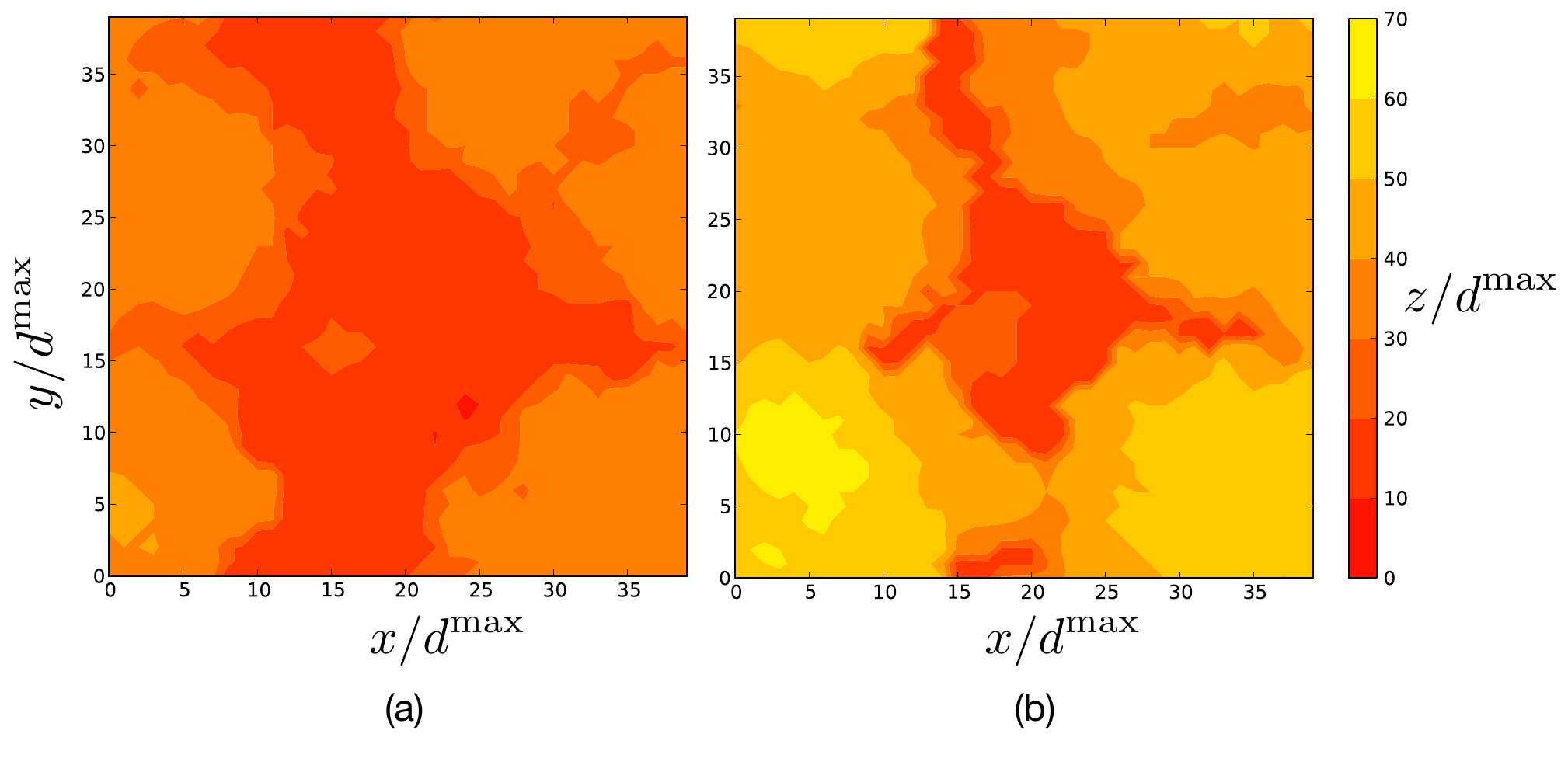}}
\caption{Contour plots showing the height $h$ as a function of horizontal coordinates $x$ and $y$ of the same system at times $t=400k_{\rm max}^{-1}$~(a) and $t=500k_{\rm max}^{-1}$~(b). The parameters are the same as in Fig.~\ref{f:snapshot}. The calibration bar on the right-hand size applies to both figures and all lengths have been scaled by the maximum particle diameter~$d^{\rm max}$.}
\label{f:idRef}
\end{figure}

\begin{figure}[htbp]
\center{\includegraphics[width=9cm]{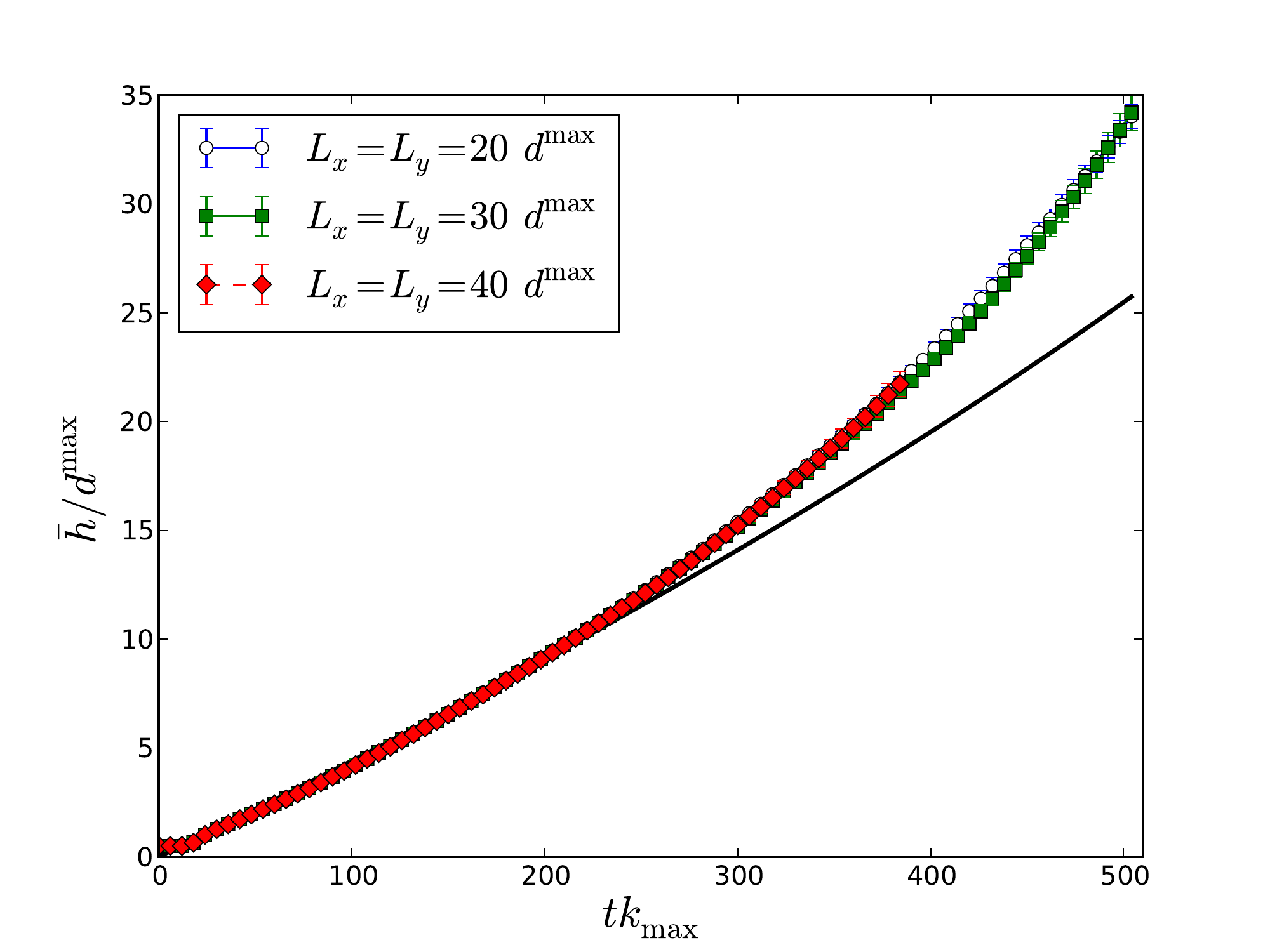}}
\caption{Mean surface height $\bar{h}$ versus time $t$ for $c_{0}=5K_{1/2}$ and no flow, $\dot{\gamma}=0$. The solid black line shows the flat film prediction from (\ref{e:thyGroThin}), which requires no fitting parameters. The height has been scaled to the threshold diameter for particle division $d^{\rm max}$, and $t$ to the base reaction rate~$k_{\rm max}$. The horizontal system sizes are given in the legend. The series for the largest system is shorter due to computational limitations.}
\label{f:height}
\end{figure}

We start with the no-flow case~$\dot{\gamma}=0$. The mean surface height $\bar{h}(t)$ is defined by
\begin{equation}
\bar{h}(t) = \frac{1}{L_{x}L_{y}}\int{\rm d}x\,\int{\rm d}y\,h(x,y)
\end{equation}
where $h(x,y)$ is the height of the surface vertically above the basal coordinates $(x,y)$ at time~$t$. This is determined using the procedure explained in Appendix~\ref{s:appData}. Contour plots of $h(x,y)$ are shown in Fig.~\ref{f:idRef}. For all parameters studied, the variation of $\bar{h}(t)$ with time showed no significant variation with the horizontal system size $L_{x}=L_{y}$. An example is given in Fig.~\ref{f:height}, where the analytical solution for a flat film (\ref{e:thyGroThin}) is also plotted. The bulk cell mass density $nm$ (where $n$ is the mean number of particles per unit volume, and $m$ the mean mass per particle) in this solution curve was measured independently, so there are no fitting parameters. Actual growth curves consistently exceed this theoretical prediction at late times. Since the degree of overshoot correlates with increasing surface roughness (as defined below; data not given), we infer this results from the omission of surface undulations in the calculations. Note that direct observation of the data confirms that $c(z<h)\ll K_{1/2}$ in all cases, as per the calculations.

%
%

%
%

\begin{figure}[htbp]
\center{\includegraphics[width=9cm]{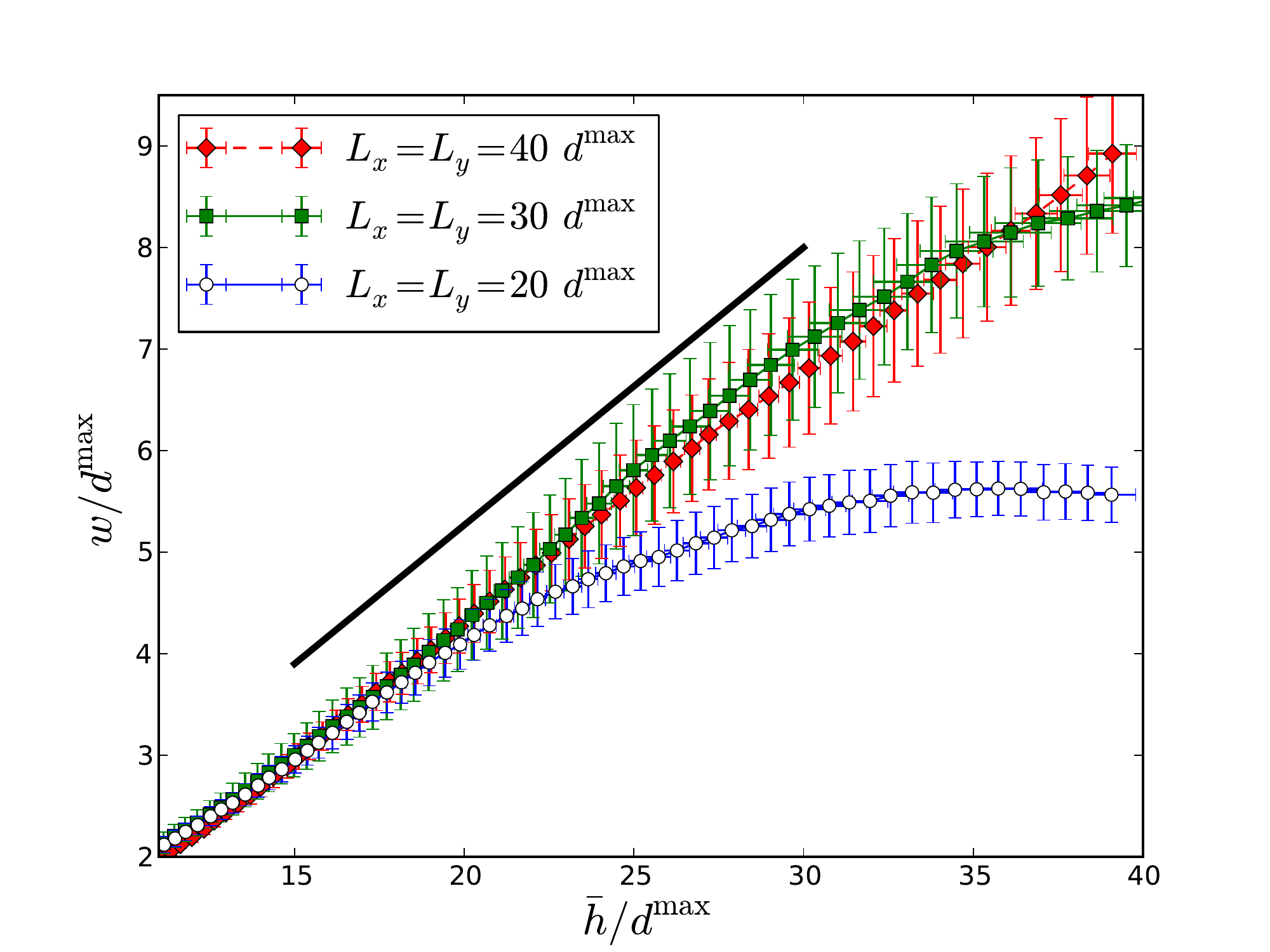}}
\caption{Surface roughness versus height for $c_{0}=10K_{1/2}$, $\dot{\gamma}=0$ and the horizontal system sizes given in the legend. For comparison, the solid black line segment has a slope of $\approx0.27$. Bars show standard error over independent runs with differently randomised initial conditions and mass redistribution after division ($n=$18, 10, 6 runs for $L_{x}=L_{y}=20d^{\rm max}$, 30$d^{\rm max}$, 40$d^{\rm max}$ resp.).}
\label{f:width}
\end{figure}

As the mean height $\bar{h}(t)$ grows non-linearly with time, unlike the canonical surface growth models where it grows at a constant rate~\cite{BarabasiBook}, we hereafter take $\bar{h}$ as a surrogate time variable to permit direct comparison with other models. We first consider the surface roughness or width $w$ defined by
\begin{equation}
w^{2}=\frac{1}{L_{x}L_{y}}\int d{\rm x}\int d{\rm y}\, \left[h(x,y)-\bar{h}\right]^{2}.
\end{equation}
The typical growth of $w$ with $\bar{h}$ is shown in Fig.~\ref{f:width}. As with fractal growth models, this increases until saturating at a maximum value that increases with system size. Unlike canonical models, where growth is sub-linear~\cite{BarabasiBook}, here the growth rate is consistent with {\em linear} scaling $w(t)\propto \bar{h}(t)$ as shown in the figure. Unfortunately the poor statistics rules out a more precise evaluation of the growth exponent.

%
%

%
%

\begin{figure}[htbp]
\center{\includegraphics[width=9cm]{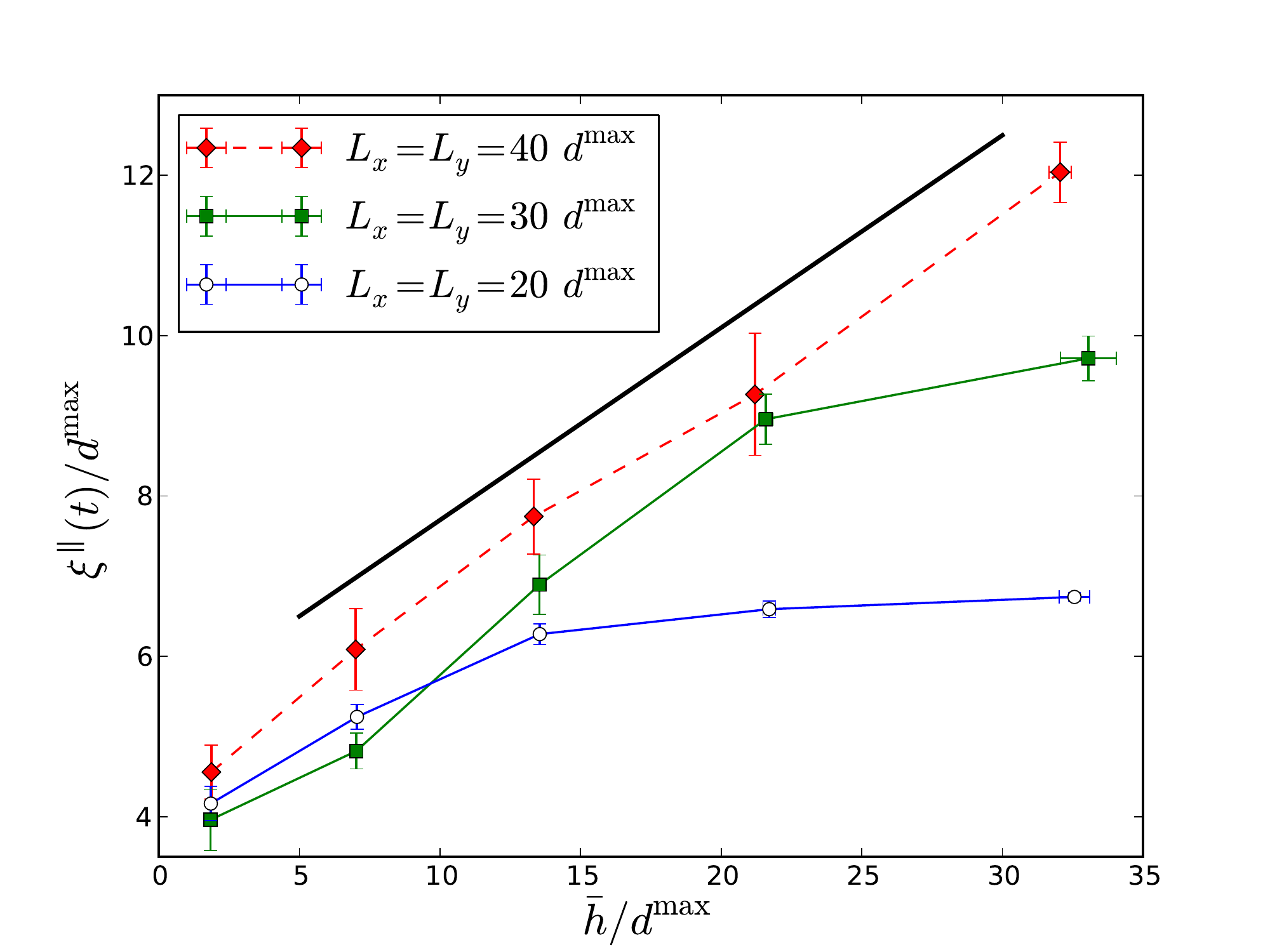}}
\caption{Growth of horizontal correlation length for the system sizes given in the legend and the same parameters as Fig.~\ref{f:width}. The straight black line segment has a slope of 0.24.}
\label{f:heightCorrn}
\end{figure}

The influence of finite system size is expected to be due to a horizontal correlation length $\xi^{\parallel}$ that grows with time, causing $w$ to saturate when $\xi^{\parallel}$ approaches $L_{x}=L_{y}$. $\xi^{\parallel}$ can be extracted from the height-height correlation function $C_{hh}(r)$,
\begin{equation}
C_{hh}(r)
=
\int{\rm d}x^{\prime}\int{\rm d}y^{\prime}\left[h(x^{\prime},y^{\prime})h(x^{\prime}+x,y^{\prime}+y)-\bar{h}^{2}\right]
\end{equation}
where $r^{2}=x^{2}+y^{2}$ and translational symmetry in the $x$-$y$ plane has been assumed. For all plots, $C_{hh}(r)$ crossed from positive at small $r$ to negative at large $r$ (data not shown); the single crossing point is identified with~$\xi^{\parallel}$. An example of the variation of $\xi^{\parallel}$ and system size is given in Fig.~\ref{f:heightCorrn}, and confirms the expected picture that $\xi^{\parallel}$ grows with time until reaching a maximum value that increases with system size. The variation of $\xi^{\parallel}$ with $\bar{h}$ before saturation is again consistent with linear growth as shown in the figure, and the data for other parameters, although noisier, are consistent with this growth law.

%
%

\begin{figure}[htbp]
\center{\includegraphics[width=9cm]{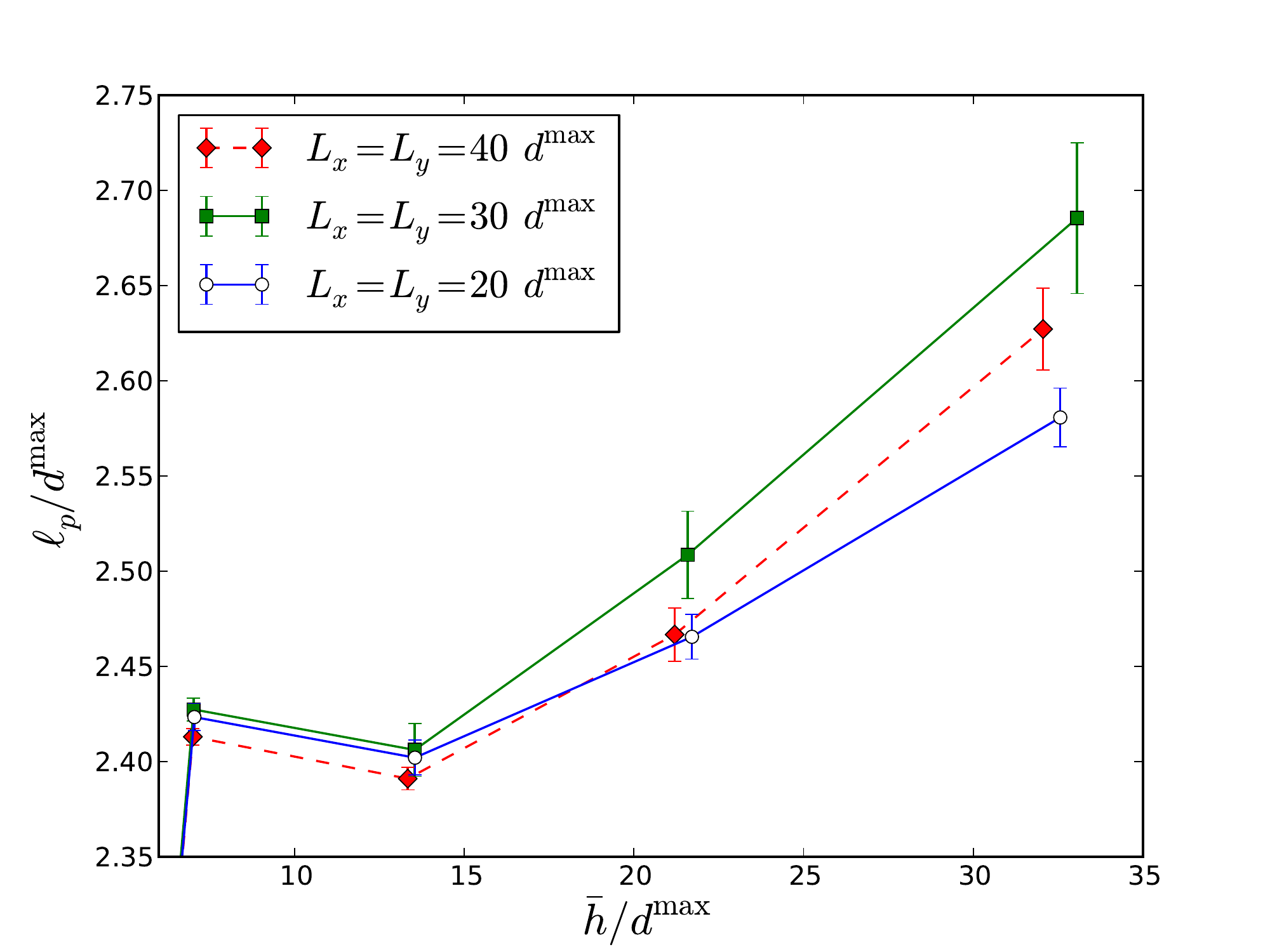}}
\caption{Depth of the active layer defined by (\ref{e:lp}) versus height for the same parameters as Fig.~\ref{f:width}, demonstrating weak variation with time. }
\label{f:surfaceActivity}
\end{figure}

In \cite{Nadell2010} it was observed that rougher films correlated with a thinner layer of actively growing particles near the upper (free) surface, as compared to a thick active layer that generated flatter films. In~\cite{Nadell2010} the thickness of the active layer was quantified by an {\em a priori} function of input parameters. Here we instead directly measure the layer thickness, and compare to the measured roughness~$w$. The active layer is defined in terms of the relative growth rate of particles $(m^{\rm c}_{i})^{-1}\partial_{t}m^{\rm c}_{i}$ measured as a function of vertical distance $\Delta z$ from the surface. Details of how this was extracted from the simulations is given in Appendix~\ref{s:appData}. The penetration depth is then
\begin{equation}
\ell_{p} =
\frac{
\left. \langle (m_{i}^{\rm c})^{-1}\partial_{t}m_{i}^{\rm c} \rangle \right|_{\Delta z=0}
}
{
\left. \left(\partial_{z}\langle (m^{\rm c}_{i})^{-1}\partial_{t}m^{\rm c}_{i} \rangle\right)\right|_{\Delta z=0}
}
\label{e:lp}
\end{equation}
where the averaging $\langle\cdots\rangle$ is over all particles at the same depth $\Delta z$ below the surface, here restricted to the surface itself. The variation of $\ell_{p}$ is plotted in Fig.~\ref{f:surfaceActivity} and demonstrates weak variation with time. By contrast, the flat-film prediction (\ref{e:thyPen}) takes a value $\ell_{p}\approx1.85\,d^{\rm max}$, 20--30\% smaller than measured, and does not increase with time. Again, the likely culprit for the excess measured thickness is the inapplicability of the flat-film assumption. Note that although the theoretical prediction employs the variation of~$c(z)$ rather than the growth rate, $c$ is roughly proportional to growth for the considered parameters, so these two definitions of $\ell_{p}$ are equivalent.

%
%

\begin{figure}[htbp]
\center{\includegraphics[width=9cm]{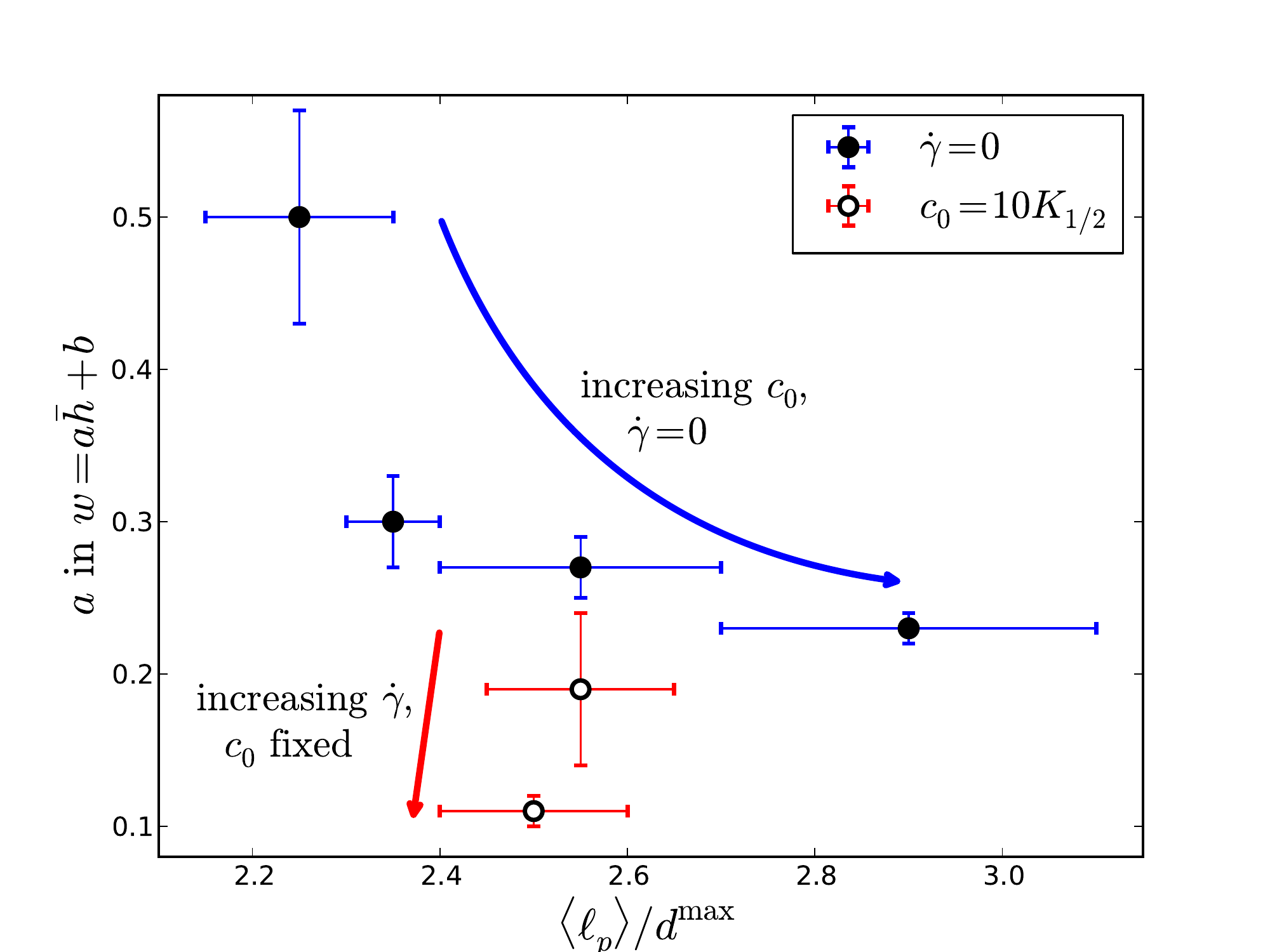}}
\caption{Surface roughness versus depth of the active layer. Closed symbols correspond to $\dot{\gamma}=0$, with increasing $c_{0}/K_{1/2}$ = 1, 5, 10 to 20 as indicated by the upper arrow. Open symbols correspond to $c_{0}=10K_{1/2}$ and increasing $\dot{\gamma}=0$, $0.072\,k_{\rm max}$ to $0.72\,k_{\rm max}$ as indicated by the lower arrow (the $\dot{\gamma}=0$ point belongs to the connecting point in the first data set).}
\label{f:roughnessVersusDepth}
\end{figure}

To correlate $\ell_{p}$ with roughness, it is convenient to reduce both time-varying quantities to single scalars. For the roughness, we focus on the linear growth regime $w=a\bar{h}+b$ and extract the slope $a$ as a measure of roughness. By choosing~$a$, which is independent of time, we have a single scalar coefficient that can be used to compare the overall increase in surface roughness for systems with different parameters (given each admits linear growth). For the active layer, we take the average of $\ell_{p}$ over the region of slow growth shown in Fig.~\ref{f:surfaceActivity}, in the understanding this is just a working definition that will weakly depend on the achievable simulation times. Plotting these two as in Fig.~\ref{f:roughnessVersusDepth} shows an inverse relationship between roughness and the depth of the active layer, confirming the finding of~\cite{Nadell2010}.

%
%

\subsection{Effect of affine flow}
\label{s:idioref}

\begin{figure}[htbp]
\centerline{\includegraphics[width=8cm]{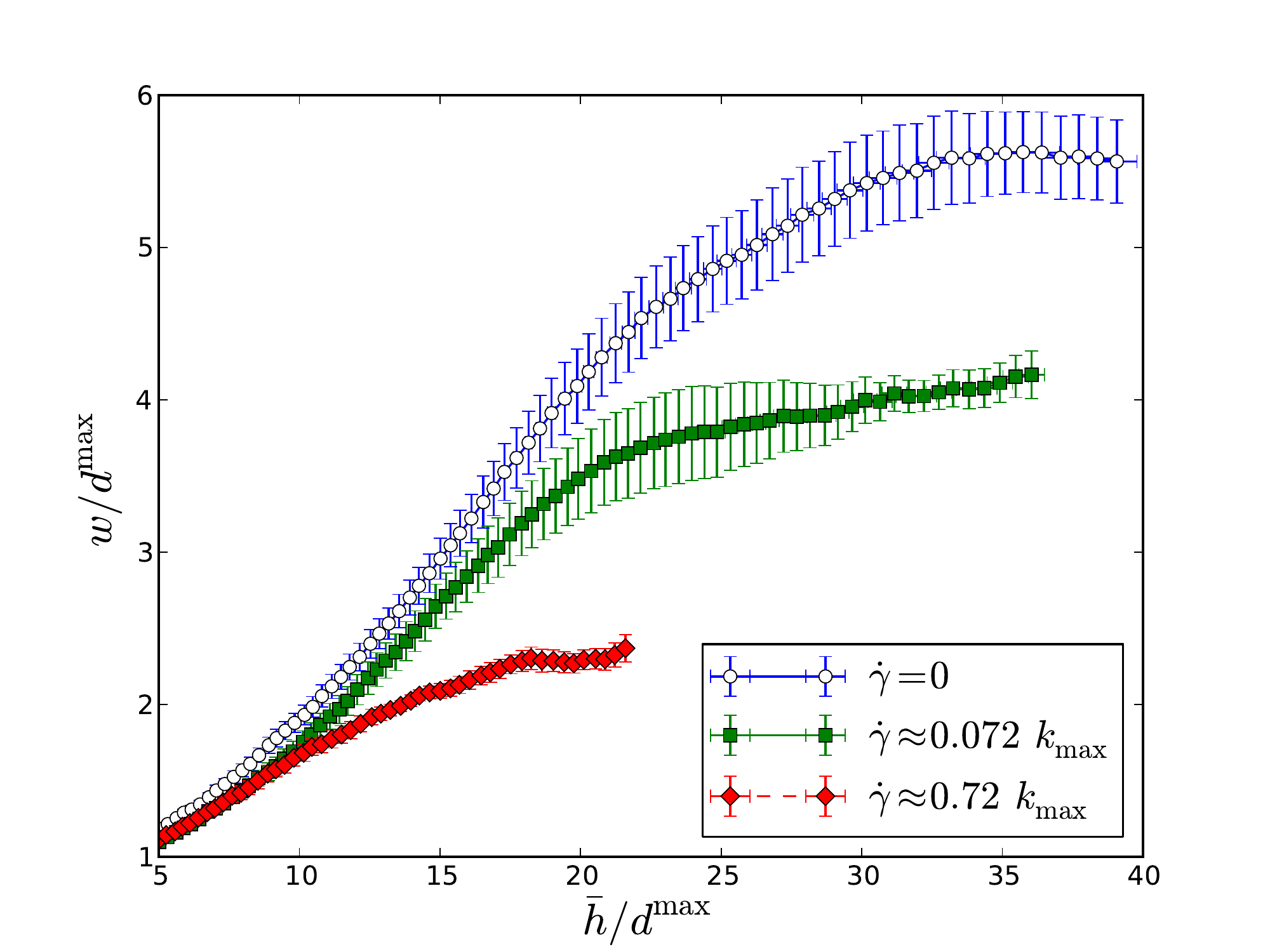}}
\caption{Growth of roughness for the shear rates given in the legend, for system size $L_{x}=L_{y}=20\,d^{\rm max}$. Note that the data sets are shorter for the fastest flow rate considered here as the mechanical stabilisation algorithm stalled for thick films, necessitating premature termination of the simulation. The solid black lines correspond to the same $\dot{\gamma}$, in the same order from top to bottom, but with a 2:1 aspect ratio in the direction of flow, {\em i.e.} $L_{x}=2L_{y}=20\sqrt{2}$ (errors bars not shown for clarity but similar to the corresponding 1:1 data).}
\label{f:flowWidth}
\end{figure}

We now turn to consider the effects of flow, $\dot{\gamma}>0$, keeping the bulk concentration fixed at~$c_{0}=10K_{1/2}$. Although the mean surface height grows at a slightly lower rate in the presence of flow, much more striking is the significant decrease in surface roughness demonstrated in Fig.~\ref{f:flowWidth}. Although the growth law remains approximately linear, the slope is noticeably reduced compared to the no-flow case. It might be postulated that the reduction in roughness is due to some change in the depth of the active layer. However, as shown in Fig.~\ref{f:roughnessVersusDepth}, flow affects the roughness but {\em not} the depth of the surface layer. Instead, this appears to be some long range interaction, as can be inferred from the data in the figure for systems with the same horizontal area $L_{x}L_{y}$ but a 2:1 aspect ratio in the direction of flow. For zero and low flow rates the curves systematically deviate from the 1:1 aspect ratio data, indicating significant system shape effects, but this modulation vanishes for the highest flow rate considered, suggesting flow reduces the range of this interaction. A likely candidate for the mechanism underlying this observation is discussed in Sec.~\ref{s:discussion}.

\begin{figure}[htbp]
\centerline{\includegraphics[width=8cm]{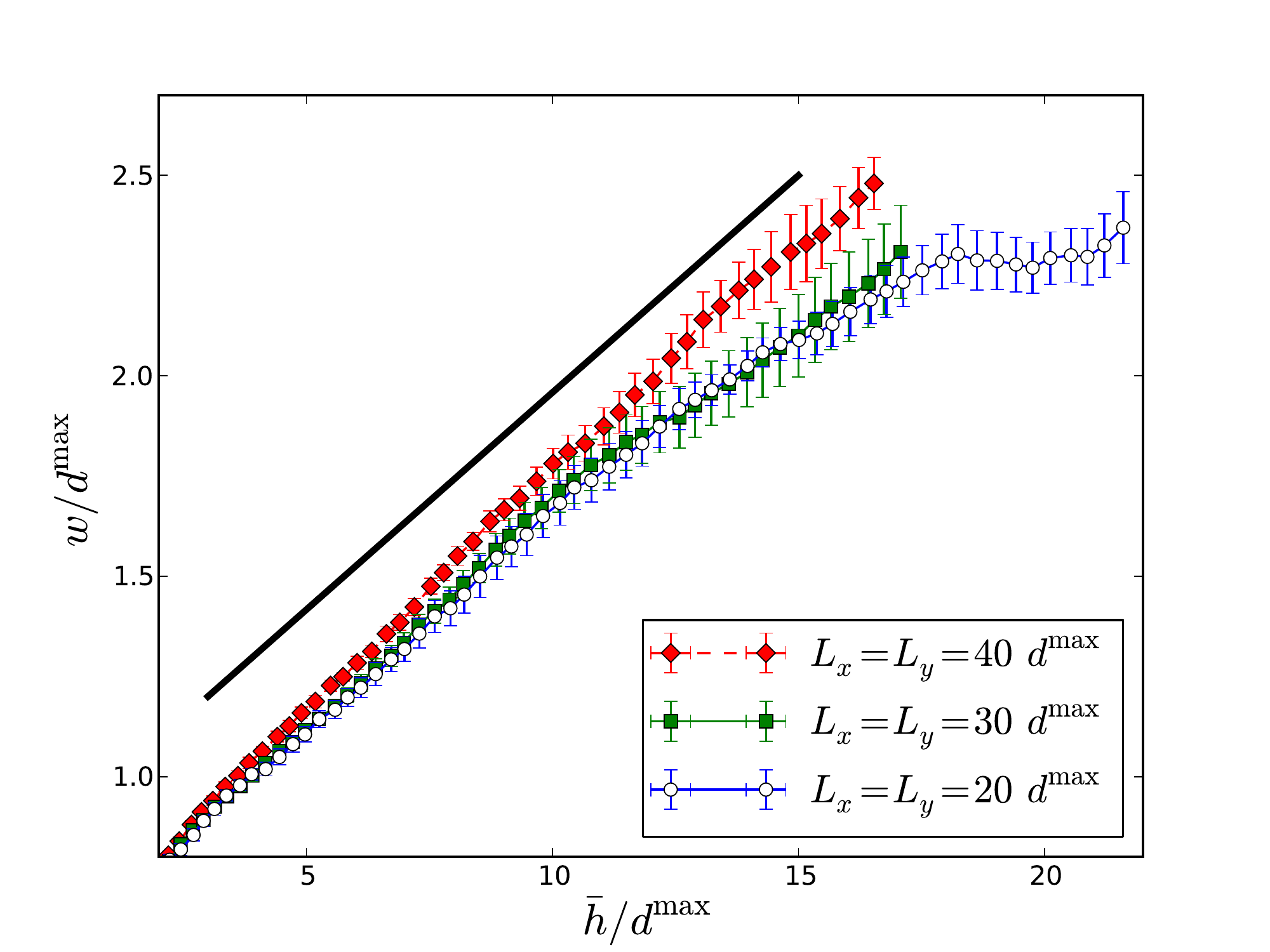}}
\caption{Variation of the growth in surface roughness with system size for $\dot{\gamma}\approx0.072\,k_{\rm max}$. Compare to the no-flow case in Fig.~\ref{f:width}.}
\label{f:flowSysSize}
\end{figure}

Systematically varying the system size reveals a mixed picture. For the highest flow rate $\dot{\gamma}\approx0.72\,k_{\rm max}$, there is no significant variation with $L_{x}=L_{y}$ as shown in Fig.~\ref{f:flowSysSize}. For the lower flow rate considered, $\dot{\gamma}\approx0.072\,k_{\rm max}$, the roughness for $L_{x}=L_{y}=20\,d^{\rm max}$ significantly exceeded that for $L_{x}=L_{y}=30\,d^{\rm max}$, taking values close to the $\dot{\gamma}=0$ case, although the statistics are poor and this observation is not definitive. This uncertainty is reflected in the large vertical error bar for this point in Fig.~\ref{f:roughnessVersusDepth}. While the reduction of roughness due to shear is clear from this figure (and outside error bars), improved statistics and a larger range of system sizes, both requiring the development of more advanced algorithms, will be required to fully clarify the picture.

A final observation relates to the mean cellular mass density, denoted $nm$ in connection with the theory of Appendix~\ref{s:appThy}. This was measured for all parameters and system sizes far from the surface, and exhibited no significant variation with system size or~$c_{0}$. It did however admit a slight but definite {\em decrease} for high flow rates, dropping roughly 5\% for the highest flow rate considered, $\dot{\gamma}\approx0.72\,k_{\rm max}$, compared to $\dot{\gamma}=0$. This is most probably an expression of {\em Reynolds' dilation}, a phenomenon common to particulate media where shear stresses generate system-spanning force chains that react against the solid surface, raising the system and lowering the mean density~\cite{Kabla2009}.

%
%

%
%

%
%
\section{Discussion}
\label{s:discussion}

Many features common to fractal growth models~\cite{BarabasiBook} have been observed in this investigation, including an algebraic increase in surface roughness in both the horizontal and vertical directions, that saturates when the horizontal  correlations $\xi^{\parallel}$ become comparable to the system size. Further evidence for fractality comes from the snapshots in Fig.~\ref{f:snapshot} and Fig.~\ref{f:idRef}, comparable results for related models~\cite{Bonachela2011}, and experiments of growing colonys (see references in~\cite{Lacasta1999,Mimura2000}). Somewhat anomalous are the growth exponents themselves, which are both consistent with linear growth, significantly faster than the sub-linear laws typically measured. An explanation based on `freezing' of surface regions might provide a simple explanation for the linear growth in~$w$, but {\em not} in~$\xi^{\parallel}$, and in any case is not consistent with direct observation of the full surface profiles which suggest no such freezing. Although the linear growth of $w$ and $\xi^{\parallel}$ suggests a {\em dynamic} exponent also equal to~1~\cite{BarabasiBook}, our statistics are too poor to permit a meaningful check of this additional exponent to confirm this relation.

It was postulated in Sec.~\ref{s:idioref} that the reduction in roughness with increasing flow rate reflects the existence of long range interactions that become shorter range for the fastest flow rate achieved, and this was supported by data for differing system aspect ratio. Here we discuss the identity of this interaction and why it may have such an effect on the surface roughness. We hypothesise that the key mechanism is non-locality deriving from the nutrient concentration field~$c({\bf x})$. In this context, it is instructive to note that the stationary Green's function ({\em i.e.} the steady solution for a point source) for~(\ref{e:rda}), in an infinite system without flow, decays with distance $|{\bf x}|$ from the source as $c^{\rm Grn}({\bf x})\propto|{\bf x}|^{-1}$~\cite{RutherfordBook}, a scale-invariant, long-range decay (in 2D the same solution does not decay at all but increases logarithmically, suggesting an even longer range effect). Non-local effects should therefore be expected. Nonetheless we have been unable to derive a simple explanation for the $w\propto\bar{h}$ growth law, and suggest that the construction of simplified models will allow larger systems to be reached and generate insight into this phenomenon. Furthermore, we cannot rule out a crossover to different scaling at late times exceeding our simulation capabilities, as in some other models with non-local surface interactions~\cite{BarabasiBook,Tang1990}. We note however that the biofilm thickness reached in our simulations, roughly $150-200\mu m$, are comparable to real biofilms, therefore our findings should be regarded as biologically relevant.

Shear flow is well known to induce waves at liquid surfaces, but can also smooth surfaces by suppressing thermal capillary waves as observed in colloidal gas-liquid interfaces~\cite{Derks2006}. This insight cannot be transferred to our athermal system, however, thus the mechanism underlying the measured reduction in roughness is not clear. Since the elastic strains in the biofilm were visibly very small, the observed smoothing is most likely due to alterations to the transport of the nutrient. It is not simply due to changes to the mean nutrient transported to the surface, however, as this would affect the depth of the active layer, which was not observed. Instead, we argue that the effect of flow on roughness can be intuitively understood as being due to the competition between diffusion, controlled by the parameter $D$, and advection due to the velocity field ${\bf v}({\bf x})=(\dot{\gamma}z,0,0)$. The effect of diffusion over advection can be quantified by the dimensionless Prandtl number $P=D/(\dot{\gamma}h^{2})$ with $h$ a characteristic height of the film. Taking $h\approx 100\mu m$ as a typical biofilm thickness, the two values of $\dot{\gamma}$ employed here, $\dot{\gamma}\approx0.072k_{\rm max}$ and $0.72k_{\rm max}$, correspond to $P\approx10$ and $P\approx1$ respectively. This confirms the relevant role of advection to our results. It does not, however, highlight the microscopic mechanism underlying the smoothing, and here again further investigation of simplified models is desirable.

This first application of the mechanical-IbM model remains deficient in two key respects. Firstly, the coupling between the fluid and the rest of the system is strictly one-way, {\em i.e.} the fluid affects the biofilm and the nutrient field, but the biofilm does not affect the flow. It can be argued this is valid for the low shear rates considered here, but will likely break down for higher rates when hydrodynamic interactions will be needed. To see this, first note that biofilms are highly porous~\cite{Zhang1994,Renslow2011,David2013}. A comparable system is therefore polymer brushes attached to a surface. Hydrodynamic simulations of polymer brushes, where the lowest strain rate considered was an order of magnitude larger than the largest considered here, have shown negligible effect on the density profile due to such low shear rates~\cite{Doyle1998}. These same simulations do show a significant reduction in fluid velocity deep within the polymeric bulk; however, since this will only affect the transport of nutrients, the concentration of which is anyway very low deep within the film, this omission will make negligible difference to biofilm growth. The second deficiency in this model is that biomass detachment due to shear stresses has not been incorporated, although this is known to partly control biofilm thickness~\cite{Chang1991,Huang2012}. There is no reason why this cannot be introduced for a future work, possibly following the particle-removal criterion of Alpkvist and Klapper~\cite{Alpkvist2007}. We note that the m-IbM model maintains the primary advantages of the IbM approach, including the relative ease of modelling multi-species films, and speculate it will become an important tool in the quantification of biofilm-flow coupling in the future.

%
%
\begin{acknowledgments}
This work was funded by the Biomedical Health Research Centre (BHRC), University of Leeds, UK.
\end{acknowledgments}

%
%
\appendix

%
%
\section{Flat-film theory}
\label{s:appThy}

For flat films with uniform thickness~$h(t)$, it is possible to write down analytically tractable equations by approximating the biofilm as a continuous body. The discrete reactions $r_{i}$ in (\ref{e:monod}) are replaced by the continuous field $r(z)$, which is proportional to the number density of cells per unit volume, $n$, and the mass per cell, $m$, both of which are taken as uniform and constant. For clarity of the resulting expressions, we define $\alpha=nmk_{\rm max}$. Then $c(z)$ obeys the following one-dimensional reaction-diffusion equation,
\begin{eqnarray}
0=\partial_{t}c(z) & = & D\partial^{2}_{z}c(z) + r(z),
\label{e:thyRD_c}
\\
r(z) & = & 
\left\{
\begin{array}{c@{\quad:\quad}c}
0 & z>h(t), \\
-\alpha
\displaystyle{\frac{c(z)}{c(z)+K_{1/2}}} & z<h(t).
\end{array}
\right.
\label{e:thyRD_r}
\end{eqnarray}
Even with these simplifications, (\ref{e:thyRD_r}) is non-linear and no general analytical solution is apparent. Instead we consider limits of high and low $c$ throughout the biofilm, {\em i.e.} $c(z<h)\gg K_{1/2}$, and conversely $c(z<h)\ll K_{1/2}$, for which the non-linearity is removed and (\ref{e:thyRD_c}) can be solved. The solution for $c(z<h)\gg K_{1/2}$ is
\begin{equation}
c_{0} - c(z) = \left\{
\begin{array}{l@{\quad:\quad}c}
\vspace{5mm}
\displaystyle{\frac{\alpha h}{D}\left(L_{z}-z\right)} & z>h, \\
\displaystyle{\frac{\alpha h}{D}\left(L_{z}-\frac{h}{2}\right) - \frac{\alpha}{2D}z^{2}}
 & z<h.
\end{array}
\right.
\label{e:thyHighC}
\end{equation}
For consistency we must also have~$c(0)\gg K_{1/2}$. In the opposite limit $c(z<h)\ll K_{1/2}$, and for clarity defining $\beta^{2}=\alpha(K_{1/2}D)^{-1}$,
\begin{equation}
\frac{c(z)}{c_{0}}
=
\left\{
\begin{array}{l@{\,:\,}c}
\vspace{5mm}
1-\displaystyle{\frac{\beta(L_{z}-z)\sinh(\beta h)}{\cosh(\beta h)+\beta(L_{z}-h)\sinh(\beta h)}} & z>h, \\
\displaystyle{\frac{\cosh(\beta z)}{\cosh(\beta h)+\beta(L_{z}-h)\sinh(\beta h)}} & z<h. \\
\end{array}
\right.
\label{e:thyLowC}
\end{equation}
Here we additionally require~$c(h)\ll K_{1/2}$. For both limits, continuity of $c(z)$ and $\partial_{z}c(z)$ at $z=h(t)$ has been imposed.

For $c(z<h)\gg K_{1/2}>0$ the concentration remains significantly non-zero to the base of the film. By contrast, for $c(z<h)\ll K_{1/2}$ the concentration can become vanishingly small while still within the film. In this case we define the penetration depth $\ell_{\rm p}$ by
\begin{equation}
\ell_{\rm p} = \frac{c(h)}{\left.\partial_z c(z)\right|_{z=h}}
\label{e:thyPenDef}
\end{equation}
where continuity of the first derivative means that either of the $z<h$ or $z>h$ expressions in (\ref{e:thyLowC}) can be used, giving
\begin{equation}
\ell_{\rm p}
=
\frac{\coth(\beta h)}{\beta}
\approx
\beta^{-1}
\quad
{\rm for}\quad\beta h\gg1.
\label{e:thyPen}
\end{equation}
For thick films $\beta h\gg1$ this increases with $D$ and decreases for higher reaction rates. It is thus a length scale that determines the balance of diffusion to reaction, and plays a comparable role to the (dimensionless) Thiele modulus~\cite{Stewart1996}. Conversely for thin films $\beta h\ll1$, $\ell_{p}$ diverges, suggesting it cannot be identified with a physical length scale in the original discrete system.

The time evolution of the film thickness $h(t)$ can be determined by considering the rate of change of the total mass in the film, and maintaining the assumption of constant~$nm$. It is then straightforward to derive the following integro-differential equation from~(\ref{e:thyRD_c}) and (\ref{e:thyRD_r}),
\begin{equation}
\frac{{\rm d}h(t)}{{\rm d}t}
=
k_{\rm max}Y^{\rm c}
\int_{0}^{h(t)}{\rm d}z\, \frac{c(z)}{c(z)+K_{\rm 1/2}}
\label{e:thyGrowth}
\end{equation}
It is again simpler to remove the non-linearity in the integrand by considering limits of $c(z)$. For $c(z<h)\gg K_{1/2}$, when the nutrient penetrates throughout the entire film, (\ref{e:thyGrowth}) is readily solved to give exponential growth,
\begin{equation}
h(t) = h(0) e^{k_{\rm max}Y^{\rm c}t}
\end{equation}
For $c(z<h)\ll K_{1/2}$, (\ref{e:thyLowC}) can be used to evaluate the integral in~(\ref{e:thyGrowth}), producing the differential equation
\begin{equation}
\frac{{\rm d}h(t)}{{\rm d}t}
=
\frac{c_{0}DY^{\rm c}}{nm}
\frac{1}{\ell_{\rm p}+(L_{z}-h(t))}
\label{e:thydh}
\end{equation}
Note that there is implicit $h$-dependence in the nutrient penetration depth~$\ell_{\rm p}$ as per (\ref{e:thyPen}). Because of this, it is simplest to solve (\ref{e:thydh}) in the limits of shallow and deep nutrient penetration layers relative to the boundary layer, {\em i.e.} $\ell_{\rm p}\ll L_{z}-h$ and $\ell_{\rm p}\gg L_{z}-h$ respectively. For the former case, the solution is
\begin{equation}
h(t)
=
L_{z}-\sqrt{[L_{z}-h(0)]^{2}-\frac{2c_{0}DY^{\rm c}}{nm}t}
\label{e:thyGroThin}
\end{equation}
This expression predicts the film reaches $L_{z}$ at a finite time $[L_{z}-h(0)]^{2}nm/(2c_{0}DY^{\rm c})$, but the assumption $\ell_{p}\ll L_{z}-h(t)$ will break-down before this happens. The corresponding solution for the deep penetration depth limit $\ell_{\rm p}\gg L_{z}-h$ is
\begin{equation}
h(t)=\beta^{-1}{\rm arsinh}\left\{
\sinh[\beta h(0)]
\exp\left(
\frac{c_{0}k_{\rm max}Y^{\rm c}}{K_{1/2}}t
\right)
\right\}
\end{equation}
Finally, note that the crossover between shallow and deep penetration can be expressed in terms of the dimensionless ratio $\ell_{\rm p}/(L_{z}-h)$, which (for $\ell_{p}\approx\beta^{-1}$) gives a similar quantity to the $\delta$ employed in~\cite{Nadell2010}.

%
%

\section{Data analysis of the surface}
\label{s:appData}

To determine the moments of the height distribution it is first necessary to identify the surface. To do this, the system box was partitioned into a cubic lattice in which each block has dimensions $d^{\rm max}\times d^{\rm max}\times d^{\rm max}$. Every lattice block with one or more particle centres ${\bf x}_{i}$ contained within it was marked as occupied; all others are marked vacant. Lattice blocks on the base, {\em i.e.} in the plane $z=0$, are labelled $k$ and $l$ in the $x$ and $y$-directions respectively. The height $h_{kl}$ of the film above each base block is defined as the midpoint of the highest occupied block vertically above it. Note that this definition ignores overhangs, but as these were rarely observed they should only represent a small correction to our basic findings. Moments of $h_{jk}$ were calculated as per any discrete distribution. For the spatial correlations in height, the horizontal distance $r$ between midpoints of base lattice blocks were used, incorporating the periodic boundary conditions in the horizontal directions.

Metabolic activity as a function of distance from the surface was measured using the same lattice. In this case, the mean relative growth rate $m^{-1}\partial_{t}m$ of all particles in each lattice block were calculated and assigned to that block. This was output as a function of distance from the highest occupied site in the same column~$(k,l)$, so a depth of 0 corresponds to the growth rate of particles in the highest occupied block, $d^{\rm max}$ to the block directly beneath it, {\em etc.}


\end{document}